# Effects of hydrostatic pressure on the magnetic susceptibility of ruthenium oxide $Sr_3Ru_2O_7$: Evidence for pressure-enhanced antiferromagnetic instability

Yuri V. Sushko, Bruno DeHarak, Gang Cao, G. Shaw, D. K. Powell, and J. W. Brill

*Department of Physics & Astronomy, University of Kentucky, Lexington, KY 40506*

(March 7, 2003)

Hydrostatic pressure effects on the temperature- and magnetic field dependencies of the in-plane and out-of-plane magnetization of the bi-layered perovskite $Sr_3Ru_2O_7$ have been studied by SQUID magnetometer measurements under a hydrostatic helium-gas pressure. The anomalously enhanced low-temperature value of the paramagnetic susceptibility has been found to systematically decrease with increasing pressure. The effect is accompanied by an increase of the temperature $T_{max}$ of a pronounced peak of susceptibility. Thus, magnetization measurements under hydrostatic pressure reveal that the lattice contraction in the structure of $Sr_3Ru_2O_7$ promotes antiferromagnetism and not ferromagnetism, contrary to the previous beliefs. The effects can be explained by the enhancement of the inter-bi-layer antiferromagnetic spin coupling, driven by the shortening of the superexchange path, and suppression, due to the band-broadening effect, of competing itinerant ferromagnetic correlations.

PACS numbers: 75.30.Kz, 75.40.Cx, 62.50.+p

## I. INTRODUCTION

Close proximity of superconductivity to a magnetically ordered state is emerging as a universal property of virtually all novel classes of superconductors. Examples include not only recently discovered "ferromagnetic" superconductors[1,2], but also long-time and extensively studied heavy-fermion compounds, organic charge-transfer salts, and rare-earth nickel boride carbides, in which superconductivity competes with antiferromagnetism, and, of course, the much celebrated high-$T_c$ cuprates, with an antiferromagnetic Mott insulator ground state in the non-doped parent compounds and spin fluctuation-mediated ($d$-wave) superconductivity in the optimally doped ones.

The only known non-cuprate superconductor with a perovskite structure, the ruthenium oxide $Sr_2RuO_4$, is by no means an exception.[3] Triggered initially by an observation of a close similarity of crystal and electronic structure between $Sr_2RuO_4$ and itinerant ferromagnet[4] $SrRuO_3$, the discussions on magnetism in the ruthenates family and its effects on superconducting pairing mechanism in $Sr_2RuO_4$ continues to be a hot topic in the current literature.

The dominant viewpoint treats $Sr_2RuO_4$ as an unconventional spin-triplet superconductor, and indeed numerous theoretical results and large number of experimental data suggest that pairing in $Sr_2RuO_4$ is mediated by ferromagnetic spin correlations[5-10]. However, a new and rather intriguing aspect emerged recently when an antiferromagnetic Mott insulator state was reported[11-13] in the closely related layered perovskite, $Ca_2RuO_4$. This, and also a theoretical prediction[14] and experimental observation[15] of antiferromagnetic spin fluctuations in $Sr_2RuO_4$ (apparently driven by the Fermi-surface nesting), suggest that a tendency to antiferromagnetism might be a feature common to all ruthenates with a perovskite-derived structure, just as in cuprates. If so, an antiferromagnetic spin-fluctuation mediated pairing might be also a possible mechanism of superconductivity in $Sr_2RuO_4$.[14,16]

In such a context, valuable insight into the intriguing relationship between superconductivity and magnetism can be obtained via detailed investigation of the magnetic properties of the compound that is most closely related (structurally and electronically) to $Sr_2RuO_4$, namely a two-dimensional metal $Sr_3Ru_2O_7$ with a bi-layer perovskite structure.

Both $Sr_2RuO_4$ and $Sr_3Ru_2O_7$ are members of a wide class of materials known as Ruddlesden-Popper series with a general chemical formula $A_{n+1}B_nO_{3n+1}$. In this series, a two-dimensional network (layer) of corner-shared octahedra $BO_6$ serves as the main structural element so that each layer couples to the adjacent ones either directly (cubic perovskites) or, as in layered and multi-layered perovskites, via an insulating rock-salt layer(s) AO. The end-point members of the series, the three-dimensional cubic perovskite $SrRuO_3$ and the 2-dimensional single-layered (n = 1) $Sr_2RuO_4$ show clear signs of a thermodynamic transition into an ordered state at finite temperatures: $SrRuO_3$ orders ferromagnetically[4] below $T_C$ of ~160K, and $Sr_2RuO_4$ becomes a superconductor[3] with $T_c$ = 1.1K. What has been learned so far about the magnetic properties of the intermediate double-layered (n = 2) perovskite $Sr_3Ru_2O_7$ turns out to be a rather complex and even controversial picture[17-26]. On the one side, a tendency to ferromagnetism is apparent, as exhibited by a strongly enhanced paramagnetic susceptibility with anomalously large[18-20] Wilson ratio (which is the dimensionless ratio of the low-temperature spin susceptibility to the electronic specific heat coefficient) $R_W$ > 10, and also by strong 2-D ferromagnetic spin correlations seen in neutron experiments.[21] On the other side, a strong antiferromagnetic instability is also present as evidenced[17-19] by the negative Weiss temperature, and a pronounced cusp in the temperature dependence of magnetic susceptibility at around $T_{max}$= 16 K. The latter anomaly is also accompanied by anomalies in specific heat[18,19], resistivity[17,19,22], Hall coefficient[22], and even by a sign reversal of the magnetoresistance[22] at the same temperature. However, despite the fact that magnetic and transport properties show clear signs of both the antiferromagnetic and the ferromagnetic instabilities, neither the specific heat[18,19] nor neutron scattering experiments[21,23] yield solid evidence for the thermodynamic phase transition into a magnetically ordered state. Apparently, neither of the two competing magnetic instabilities is strong enough to overcome the influence of the antagonist and to result in a long-range magnetic structure. Thus, it is generally acknowledged that $Sr_3Ru_2O_7$ remains in a paramagnetic state down to the lowest temperatures, although a signature of a metamagnetic quantum critical point has been reported under applied magnetic fields.[24,25]

Concerning an antiferro- vs. ferro-antagonism within the spin system of $Sr_3Ru_2O_7$, a competition between the ferromagnetic (itinerant) spin correlations in the highly conducting *ab* plane and the antiferromagnetic (superexchange) coupling along the poorly-conducting out-of-plane direction definitely is a key aspect. Thus, pressure *P*, the role of which is detrimental to the effects of reduced dimensionality, should be an important external parameter in investigating the nature of magnetism in two-dimensional ruthenates, $Sr_3Ru_2O_7$ included.

Previously, there were several attempts to study the effect of pressure on the magnetic properties of $Sr_3Ru_2O_7$. In 1998, Ikeda et al. showed[18], that introducing the smaller Ca ion into the sites of the Sr ion resulted in nearly linear contraction of the unit cell volume within the series $Sr_{3-x}Ca_xRu_2O_7$ for the wide range of Ca concentration 0 < *x* <2. They also found that such a contraction was accompanied by the overall increase of the temperature of the susceptibility maximum $T_{max}$, and, simultaneously, by the sign reversal of the Weiss temperature from negative to positive when *x* approached *x* = 1 from below. Noticeably, it is very difficult to find an appropriate interpretation of these two findings together for they seem to indicate that the lattice contraction causes the strength of the antiferromagnetic interactions to increase (as suggested by the increase of $T_{max}$) and at the same time to decrease (as suggested by the suppression of the negative Weiss temperature). Apparently, besides the purely geometric effect of a lattice contraction, the partial Ca for Sr substitution introduces additional effects associated with lattice defects, crystalline and chemical disorder, and even impurity phases. (The authors of Ref. [18] acknowledged that at least for some ranges of $x$ they results were affected by the presence of an impurity phase $SrRuO_3$, which is ferromagnetic.)

In the more recent study of effect of pressure on magnetism of $Sr_3Ru_2O_7$, Ikeda and coworkers reported[19] that an external pressure of ~10 kbar induces a dramatic increase of the field-cooled magnetization below 70K accompanied by a hysteresis in the isothermal magnetization $M(H)$ at low temperatures; the results were claimed to evidence that under hydrostatic pressure $Sr_3Ru_2O_7$ undergoes transition from a paramagnetic into a ferromagnetic state. One could argue however, regarding the interpretation of the high-pressure data in Ref. [19], that such features as a pronounced maximum in the ZFC branch of the $M(T)$ dependence and an increase of $M$ with increasing $H$ without saturation, would rather imply an antiferromagnetic state (with the weak ferromagnetic component that typically arises from the canting of antiferromagnetic sublattices) but not a ferromagnetic one. Such a controversy about interpretation, as well as the fact that magnetization data for pressures other than 1 bar and 10 kbar were not reported and are still lacking, strongly suggest that systematic and detailed pressure studies of magnetic properties of $Sr_3Ru_2O_7$ must be conducted. This paper presents the results of such a study.

## II. EXPERIMENT

Single crystals of $Sr_3Ru_2O_7$ were grown in Pt crucibles using self-flux techniques from off-stoichiometric quantities of $RuO_2$, $SrCO_3$, and $SrCl_2$. These mixtures were heated to 1480°C in Pt crucibles, fired for 20 hours, cooled at 2°C/hour to 1370°C. The obtained single crystals were characterized by powder x-ray diffraction and transmission electron microscopy (TEM) using a JEOL 2010 microscope operated at 200 kV. The composition of the crystals was examined by energy-dispersive x-ray (EDX) spectroscopy, confirming the ratio of Sr:Ru to be 3:2. All crystals used in this study were as-grown. They do not contain impurity phases, such as $SrRuO_3$ which were observed in polycrystalline samples[18] or $Sr_4Ru_3O_{10}$ that were present at flux-grown samples studied earlier[26,27]. We also confirmed the excellent quality of the crystals used in the high pressure experiments by measuring the temperature dependencies of its magnetic susceptibility, resistivity, and heat capacity.

The specific heat measurements were made on an 8 mg crystal using ac calorimetry, as described in detail in Ref. [28]. The sample was heated with light chopped at 9 Hz. Since the absorbed power was not known, the specific heat was normalized to the published results[19] at T = 27 K.

The electrical resistivity was measured with an ac four-probe method. The hydrostatic pressure of up to 13 kbar was applied using the piston-cylinder apparatus made of BeCu.

The temperature- and field dependencies of the magnetization under hydrostatic pressure were measured with a SQUID magnetometer incorporated into a helium-gas high-pressure system. The chief advantages of using helium as a pressure medium are that the pressure is truly hydrostatic, and that the value of pressure is easily controlled (and can be tuned) during the cooling/warming cycles. In our experiments, a single crystal sample was placed inside a specially designed long and slim pressure cell which was connected via a long capillary tubing to a U11 (Unipress ™) gas-compressor system. The pressure cell was inserted into the sample chamber of the commercial MPMS-5 (Quantum Design™) SQUID magnetometer with the high-pressure capillary tube playing the role of the MPMS system's sample transport rod. Both the pressure cell and the capillary were made of BeCu. The dimensions of the cell (length of 180 mm, outer diameter of 8.6 mm, and inner diameter of 3.6 mm) were identical to those of the ice-bomb type pressure apparatus that previously has been used successfully in measurements of rather week magnetic signals with a commercial Quantum Design SQUID magnetometer.[29,30] As a sample holder we employed a 130 mm long tube made of polyimid (3 mm in diameter and 40 μm in thickness) with two tiny pieces of a cotton cigarette filter held by friction inside the tube. The sample itself was sandwiched between these two cotton slabs in a proper orientation with respect to the direction of the applied magnetic field. The pressure of helium gas inside the cell was monitored by the resistivity of a manganin gauge located at the output of the last stage of compressor. The low temperature value of pressure was also controlled by measuring the value of superconducting $T_c$ for a tiny tip of high-purity tin placed inside the sample holder *immediately next* to the crystal under investigation.

**III. RESULTS AND DISCUSSIONS**

Fig. 1 shows the temperature dependence of a static magnetic susceptibility $\chi = M/H$ measured under a slightly elevated external pressure of 130 bar with the field of 1 kOe applied in the *ab* plane. The linear dependence of the inverse susceptibility, $1/\chi$, vs. temperature for T > 180 confirms Curie-Weiss type of behavior associated with the localized $Ru^{4+}$ moments and a *negative* Weiss temperature $\Theta_{CW}$ = -40K, in full agreement with the results reported by Ikeda and coworkers for the floating-zone (FZ) grown crystals. A pronounced dip in $1/\chi$ vs. *T* curve corresponding to a maximum in the $\chi(T)$ dependence is observed at $T = T_{max} = 16K$, again in an excellent agreement with the data reported for the FZ grown crystals.

Shown in Fig. 2 are the specific heat data plotted as $C_p/T$ vs. $T^2$. Although, as already commented by other workers[18-20], the temperature dependence of the specific heat of $Sr_3Ru_2O_7$ does not show any evident signs of the thermodynamic transition, the broad hump-like feature that offsets at the same characteristic temperature $T_{max}$ is clearly seen. In fact, the temperature dependence of the specific heat we measured is the same as that reported by Ikeda *et al.*[19] for the FZ grown crystal, with a Sommerfeld constant $\gamma \approx 100$ mJ / ($K^2$ Ru mol), derived from our $C_p/T$ vs $T^2$ data, significantly larger than the value of 63 mJ / ($K^2$ Ru mol) found for polycrystalline samples.[17, 18] We thus conclude that the crystals which we use for the high-pressure measurements exhibit the behavior typical for the high-quality crystals of a pure single-phase $Sr_3Ru_2O_7$ material.

The temperature dependence of magnetization under various pressures for two single crystal and two different field orientations is shown in Fig. 3. Fig.3a exhibits the results for the magnetic field ***H*** applied in the *ab* plane. The data for ***H*** perpendicular to the *ab* plane are shown in Fig. 3b. For both orientations the *M(T)* dependence under an applied pressure is found to be qualitatively the same as at ambient pressure. Namely, in a pattern typical of a localized-moment antiferromagnet, the magnetization *M* initially grows with decreasing *T*, shows a pronounced cusp at a characteristic temperature $T_{max}$ and drops rapidly upon further cooling at T< $T_{max}$. Moreover, similar to the ambient pressure data, no hysteresis between ZFC and FC branches of the *M(T)* dependence has been observed. The isothermal magnetization measurements conducted as a function of an applied magnetic field ***H*** // c at temperatures *T* = 1.8 and 5K and applied pressure of 8.15 kbar also reproduced the ambient pressure behavior. Furthermore, no evidence for a pressure-induced phase transition were found in the complementary measurements of resistivity that we performed in the range of pressures up to 12.1 kbar-- well exceeding the maximum pressures of both ours and Ikeda's et al. magnetic measurements. The temperature dependence of resistivity of the crystal #2 at two different pressures is exhibited by an inset of Fig. 3b. Clearly, the $R(T)$ behavior seen at P=1 bar and P= 12.1 kbar is qualitatively the same (and also not different from $R(T)$ observed at several intermediate pressures, although those data have been omitted for they lie sufficiently close to confuse the plot).

We thus conclude that the pressure-induced changeover from paramagnetism to ferromagnetism reported in Ref. [19] is not observed under hydrostatic helium-gas pressure. Quite to the contrary, a suppression of a ferromagnetic instability and an enhancement of antiferromagnetic instability have been revealed in our magnetization measurements. Indeed, the raw data of Fig.3 show that the position of a peak of the $M$ vs. $T$ dependence shifts with pressure to higher temperatures. Simultaneously, the low-temperature value of magnetic susceptibility decreases strongly with pressure. These effects are summarized in Fig.4, where two parameters representing the two competing magnetic instabilities are plotted as a function of pressure: the temperature $T_{max}$ at which the susceptibility starts to drop in a fashion typical for an antiferromagnet and the peak value of the in-plane magnetic susceptibility, $\chi_{ab}^* = \chi_{ab}(T_{max})$. Apparently, the temperature $T_{max}$, being a characteristic temperature of a short-range antiferromagnetic ordering provides the measure for a strength of the antiferromagnetic coupling constant $J_{AF}$ (as the Neel temperature $T_N$ does) whereas $\chi_{ab}^*$ can serve as a good estimate for a contribution of the in-plane ferromagnetic correlations ($J_{FM}$) into the strongly enhanced paramagnetic susceptibility of $Sr_3Ru_2O_7$. Noticeable pressure dependence exhibited by each of these parameters and also by their product $\chi_{ab}^* T_{max}$ contrasts with the case of simple paramagnetism, for which $\chi T$ must be constant. In particular, the dramatic drop of $\chi_{ab}^*$ under pressure (a 30% reduction is observed at P= 8.5 kbar) suggests that an applied pressure drives $Sr_3Ru_2O_7$ rapidly away from the nearly ferromagnetic state observed at ambient pressure. Simultaneously, a positive baric coefficient $d(lnT_{max})/dP = + 2\%$ /kbar provides a clear sign of the positive effect of pressure on the antiferromagnetic spin coupling.

To offer an explanation to the observed pressure effects we would like to refer to the theoretical results of Singh and Mazin.[31] In their density functional calculations of the electronic structure of $Sr_3Ru_2O_7$ (for the orthorhombic structure based on the recent neutron diffraction data[32]) the following three stable magnetic solutions were obtained: (a) the antiferromagnetic state where the Ru ions in a layer are ferromagnetically aligned, but the layers are coupled antiferromagnetically, (b) the ferromagnetic state in which both the in-plane and out-of-plane interactions are ferromagnetic, and (c) the antiferromagnetic state where bi-layers are ferromagnetic but stacked antiferromagnetically.

The experimental picture of the magnetism in $Sr_3Ru_2O_7$ with both the ferromagnetic and antiferromagnetic correlations present could be consistent with the spin arrangement leading to the structures of either type (a) or type (c) of the above classification, provided that out-of-plane superexchange coupling is rather underdeveloped and thus prevents a long-ranged 3-dimensional magnetic order from happening. The latter assumption is not difficult at all to justify in a case of inter-bi-layer spin-spin interaction through the rock-salt layers which indeed is rather weak superexchange via a long path containing two oxygen ions and unfavorable bond angles. Within such a scenario of competing itinerant ferromagnetic (in-plane) and superexchange antiferromagnetic (out-of-plane) interactions, an isotropic lattice compression under hydrostatic pressure should result in increased $J_{AF}$ (shorter superexchange path) and decreased $J_{FM}$ (the negative $dJ_{FM}/dP$ is expected in itinerant ferromagnets due to the band-broadening effect of pressure) - the two effects which are simultaneously observed in our experiments.

Interestingly enough, the observed suppression of the ferromagnetic effects under hydrostatic pressure also provides a clue for an explanation of the effect of enhanced ferromagnetic interactions revealed in the magnetization measurements of Ref. [19]. Since these measurements were conducted with a piston-cylinder pressure technique, one should expect their results to be affected by the presence of a strong non-hydrostatic-pressure component. Indeed, significant (as large as 30%-40%) and *uncontrollable* changes of pressure upon cooling due to solidification of a pressure–transmitting liquid and differential thermal contraction of the body of the cell and its interior elements (such as the sample holder, the sample itself, and the pressure medium) is a well known drawback of the clamped-cell method in general.[33] The apparatus[34] used in the experiments of Ikeda and coworkers differed from a typical piston-cylinder pressure cell in that it had a long body (~200mm) and small inner diameter (<3mm). Moreover, in order to minimize the piston displacement during the pressurization process, two long quartz rods were placed inside, serving as spacers.[34] The very important detail of this design is that the sample under investigation was literally sandwiched between the faces of these quartz roads. When such a pressure cell is cooled down in an MPMS cryostat, particularly severe effects of inhomogeneity of pressure affecting the crystal inside the cell should arise due to the combination of two factors. First, there is a substantial temperature gradient along the length of the cell, causing a substantial differential between the values of pressure in the colder lower part (solid medium) and warmer upper part (liquid medium) of the cell. As a result, the sample will be in a rather complex field of anisotropic strains and stresses instead of hydrostatic pressure conditions. The second unfavorable factor which amplifies the latter effect even more is a huge difference in the thermal contraction of the cell itself and the quartz spacers (the thermal expansion coefficient of BeCu, 17.5 x10$^{-6}$/K, is *35 times larger* than that of quartz). Contraction of the cell should be a source for an additional non-hydrostatic (longitudional) stress component acting on the sample squeezed between two incompressible quartz rods. In fact, the sample orientation for which the enhanced ferromagnetic effect was observed in Ref [19] was that with its *c*-axis parallel to the cylindrical axis of the pressure cell, favoring uniaxial stress perpendicular to the *ab*-plane of the crystal. However, as a consequence of the Poisson effect, the well known unwanted companion of uniaxial-pressure experiments[35], such stress along the *c*-axis will create a large expansion of the crystal lattice parameter in the highly conducting *ab*-plane, in turn leading to band-narrowing, an increased density of states, and, within the itinerant magnetism model, stronger ferromagnetic spin correlations.

In conclusion, using helium as a pressure-transmitting medium we investigated the hydrostatic pressure effects on the in-plane and out-of-plane magnetization of the metallic two-dimensional ruthenate $Sr_3Ru_2O_7$ by means of SQUID magnetometry. The measurements reveal monotonic suppression of the low-temperature paramagnetic susceptibility as a function of pressure and the simultaneous increase of the Neel-temperature like temperature $T_{max}$. This behavior differs from the pressure-induced changeover from paramagnetism to ferromagnetism reported in the experiments with the liquid-media piston-cylinder pressure cell.[19] Quite to the contrary, our results indicate that an applied hydrostatic pressure causes strengthening of the antiferromagnetic and suppressing of the ferromagnetic instability, effects that have a natural explanation in the Singh and Mazin model of competing in-plane itinerant ferromagnetism and out-of-plane superexchange antiferromagnetism.

This research was supported in part by the National Science Foundation, grants #DMR-9731257 and DMR-0100572.

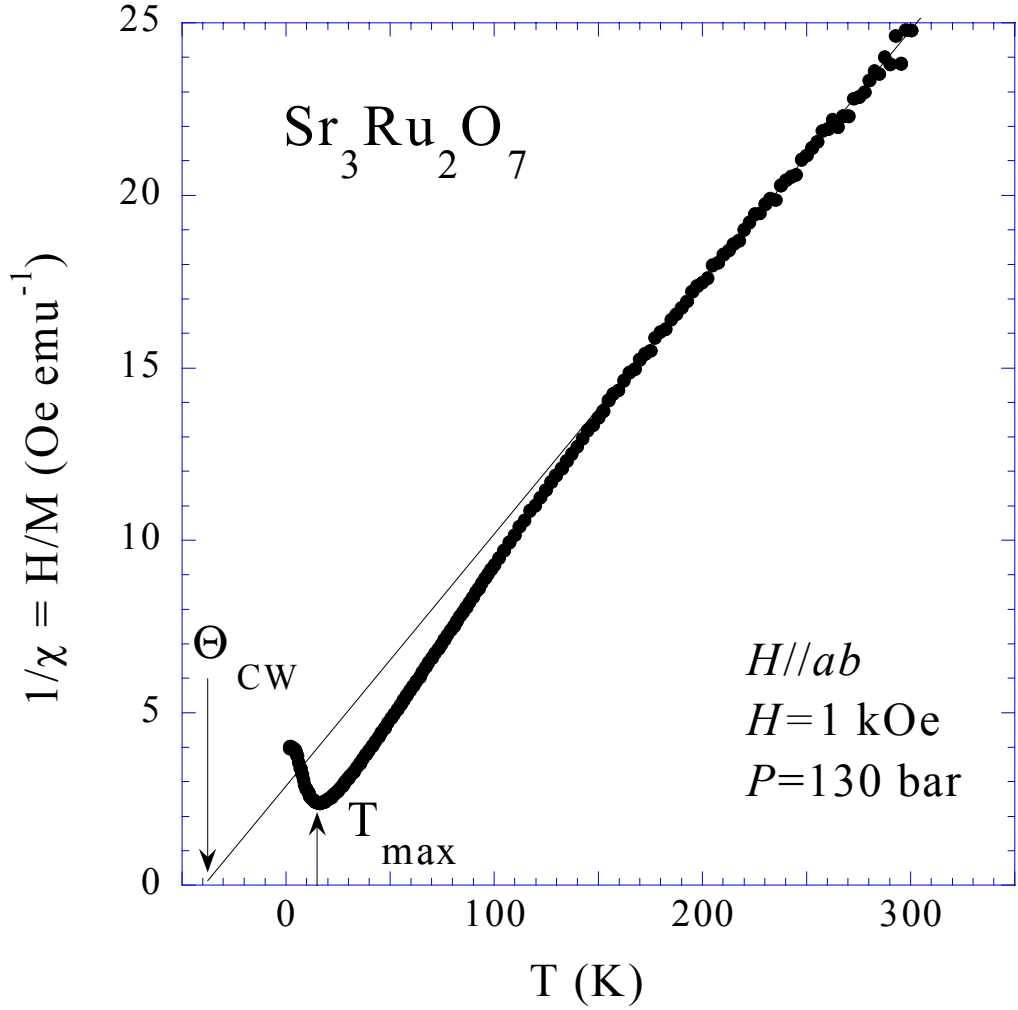

FIG. 1. The reciprocal magnetic susceptibility of $Sr_3Ru_2O_7$ under pressure of 0.13 kbar. The thin line illustrates how the value of the Weiss temperature $\Theta_{CW}$ was obtained.

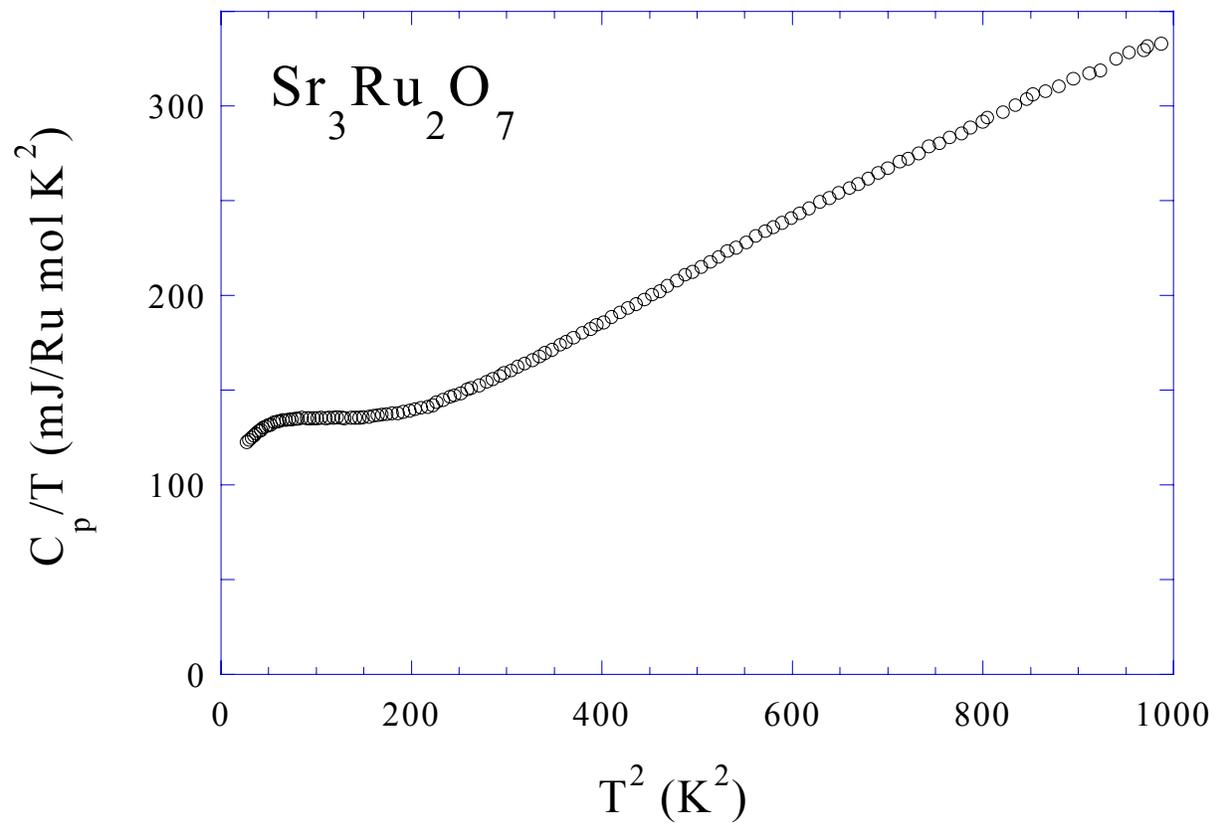

FIG. 2. The specific heat divided by temperature $C_p/T$ of a single crystal of $Sr_3Ru_2O_7$ at ambient pressure.

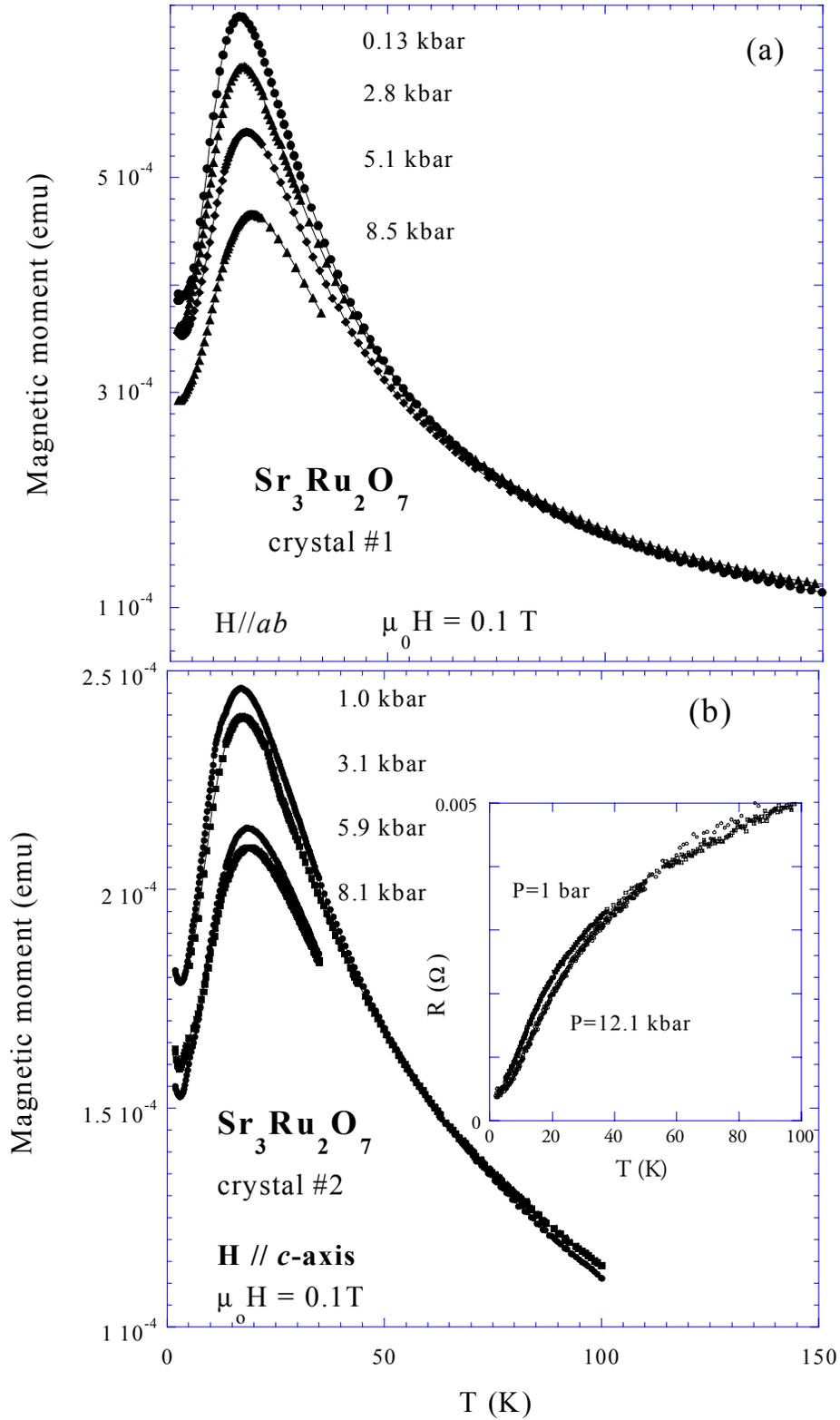

FIG. 3. Magnetization as a function of temperature of a single crystal of $Sr_3Ru_2O_7$ at different pressures $P$ and two different field orientations, $H // ab$ (Fig. 3a) and $H//c$ (Fig. 3b). For each field orientation the height of magnetization maximum decreases with increase in $P$. The data for two crystals are presented. The inset shows the temperature dependence of the in-plane resistivity for two different pressures, $P=1$ bar and $P= 12.1$ kbar.

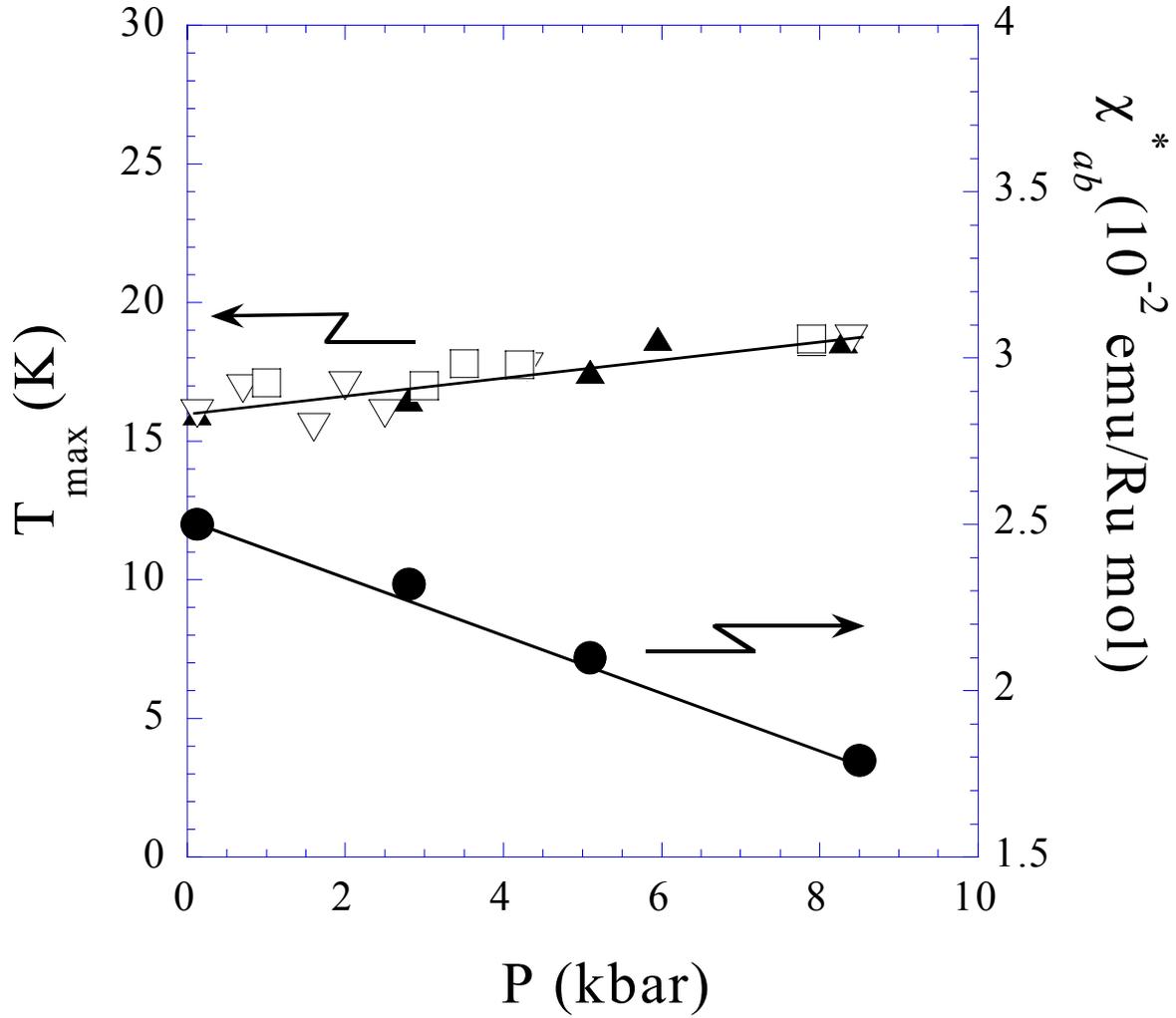

FIG. 4. The pressure dependence of two empirical parameters derived from the data of Fig. 3., the temperature $T_{max}$ and the peak value of the in-plane susceptibility $\chi_{ab}^{*}$ (see text). The values of $T_{max}$ for two crystals (triangles for the crystal #1 and squares for the crystal #2) and two field orientations (filled symbols for H//ab and empty symbols for H//c) are presented. Lines are guides for the eye.